\documentclass[conference]{IEEEtran}
\usepackage{lmodern}
\usepackage{amssymb,amsmath}
\usepackage{ifxetex,ifluatex}
\usepackage{fixltx2e} 
\ifnum 0\ifxetex 1\fi\ifluatex 1\fi=0 
  \usepackage[T1]{fontenc}
  \usepackage[utf8]{inputenc}
\else 
  \ifxetex
    \usepackage{mathspec}
  \else
    \usepackage{fontspec}
  \fi
  \defaultfontfeatures{Ligatures=TeX,Scale=MatchLowercase}
\fi
\IfFileExists{upquote.sty}{\usepackage{upquote}}{}
\IfFileExists{microtype.sty}{%
\usepackage{microtype}
\UseMicrotypeSet[protrusion]{basicmath} 
}{}
\usepackage{xcolor}
\usepackage[bookmarksnumbered,unicode]{hyperref}
\hypersetup{
            pdftitle={OpenCL Performance Prediction using Architecture-Independent Features},
            pdfborder={0 0 0},
            breaklinks=true}
\IfFileExists{parskip.sty}{%
\usepackage{parskip}
}{
\setlength{\parindent}{0pt}
\setlength{\parskip}{6pt plus 2pt minus 1pt}
}
\providecommand{\tightlist}{%
  \setlength{\itemsep}{0pt}\setlength{\parskip}{0pt}}
\ifx\paragraph\undefined\else
\let\oldparagraph\paragraph
\renewcommand{\paragraph}[1]{\oldparagraph{#1}\mbox{}}
\fi
\ifx\subparagraph\undefined\else
\let\oldsubparagraph\subparagraph
\renewcommand{\subparagraph}[1]{\oldsubparagraph{#1}\mbox{}}
\fi

\makeatletter
\def\fps@figure{htbp}
\makeatother

\usepackage{booktabs}
\usepackage[figurename={Figure.},tablename={Table.},listfigurename={List of Figures.},listtablename={List of Tables.}]{caption}
\usepackage{tabularx}
\usepackage{cite}
\usepackage{amsmath,amssymb,amsfonts}
\usepackage{textcomp}
\usepackage{xargs}
\usepackage[colorinlistoftodos,prependcaption,textsize=small,color=yellow]{todonotes}
\usepackage{regexpatch}
\usepackage{adjustbox}
\usepackage{etoolbox}
\usepackage[noend,ruled]{algorithm2e}
\usepackage{float}
\usepackage{threeparttable}
\usepackage[binary-units=true]{siunitx}
\def\BibTeX{{\rm B\kern-.05em{\sc i\kern-.025em b}\kern-.08em T\kern-.1667em\lower.7ex\hbox{E}\kern-.125emX}}
\makeatletter
\@ifpackageloaded{subfig}{}{\usepackage{subfig}}
\@ifpackageloaded{caption}{}{\usepackage{caption}}
\AtBeginDocument{%
\renewcommand*\figurename{Figure}
\renewcommand*\tablename{Table}
}
\AtBeginDocument{%
\renewcommand*\listfigurename{List of Figures}
\renewcommand*\listtablename{List of Tables}
}
\@ifpackageloaded{float}{}{\usepackage{float}}
\floatstyle{ruled}
\@ifundefined{c@chapter}{\newfloat{codelisting}{h}{lop}}{\newfloat{codelisting}{h}{lop}[chapter]}
\floatname{codelisting}{Listing}

\makeatother

\begin{document}

\title{OpenCL Performance Prediction using Architecture-Independent Features}

\date{May 16, 2018}

\makeatletter
\let\oldlt\longtable
\let\endoldlt\endlongtable
\def\longtable{\@ifnextchar[\longtable@i \longtable@ii}
\def\longtable@i[#1]{\begin{figure}[t]
\onecolumn
\begin{minipage}{0.5\textwidth}
\oldlt[#1]
}
\def\longtable@ii{\begin{figure}[t]
\onecolumn
\begin{minipage}{0.5\textwidth}
\oldlt
}
\def\endlongtable{\endoldlt
\end{minipage}
\twocolumn
\end{figure}}
\newcommand{\removelatexerror}{\let\@latex@error\@gobble}
\xpatchcmd{\@todo}{\setkeys{todonotes}{#1}}{\setkeys{todonotes}{inline,#1}}{}{}
\newtoggle{ACM-BUILD}
\toggletrue{ACM-BUILD}
\makeatother
\author{\IEEEauthorblockN{Beau Johnston}
\IEEEauthorblockA{\textit{Research School of Computer Science} \\
\textit{Australian National University}\\
Canberra, Australia \\
beau.johnston@anu.edu.au}
\and
\IEEEauthorblockN{Greg Falzon}
\IEEEauthorblockA{\textit{School of Science and Technology} \\
\textit{University of New England}\\
Armidale, Australia \\
gfalzon2@une.edu.au}
\and
\IEEEauthorblockN{Josh Milthorpe}
\IEEEauthorblockA{\textit{Research School of Computer Science} \\
\textit{Australian National University}\\
Canberra, Australia \\
josh.milthorpe@anu.edu.au}
}

\maketitle

\begin{abstract}
OpenCL is an attractive programming model for heterogeneous high-performance computing systems, with wide support from hardware vendors and significant performance portability.
To support efficient scheduling on HPC systems it is necessary to perform accurate performance predictions for OpenCL workloads on varied compute devices, which is challenging due to diverse computation, communication and memory access characteristics which result in varying performance between devices.

The Architecture Independent Workload Characterization (AIWC) tool can be used to characterize OpenCL kernels according to a set of architecture-independent features.
This work presents a methodology where AIWC features are used to form a model capable of predicting accelerator execution times.
We used this methodology to predict execution times for a set of 37 computational kernels running on 15 different devices representing a broad range of CPU, GPU and MIC architectures.
The predictions are highly accurate, differing from the measured experimental run-times by an average of only 1.2\%, and correspond to actual execution time mispredictions of 9 $\mu$s to 1 sec according to problem size.
A previously unencountered code can be instrumented once and the AIWC metrics embedded in the kernel, to allow performance prediction across the full range of modelled devices.
The results suggest that this methodology supports correct selection of the most appropriate device for a previously unencountered code, which is highly relevant to the HPC scheduling setting.

\end{abstract}


\section{Introduction}\label{introduction}

HPC architectures are becoming increasingly heterogeneous.
This trend is increasingly apparent at the node level in supercomputer systems.
For instance, the Cori system at Lawrence Berkeley National Laboratory comprises 2,388 Cray XC40 nodes with Intel Haswell CPUs, and 9,688 Intel Xeon Phi nodes {[}1{]}.
The Summit supercomputer at Oak Ridge National Laboratory is based on the IBM Power9 CPU, which includes both NVLINK {[}2{]}, a high bandwidth interconnect between Nvidia GPUs; and CAPI, an interconnect to support FPGAs and other accelerators {[}3{]}.
Promising next-generation architectures include Fujitsu's Post-K {[}4{]}, and Cray's CS-400, which forms the platform for the Isambard supercomputer {[}5{]}.
Both architectures use ARM cores alongside other conventional accelerators, with several Intel Xeon Phi and Nvidia P100 GPUs per node.

The OpenCL programming framework is well-suited to such heterogeneous computing environments, as a single OpenCL code may be executed on multiple different device types including most CPU, GPU and FPGA devices.
Predicting the performance of a particular application on a given device is challenging due to complex interactions between the computational requirements of the code and the capabilities of the target device.
Certain classes of application are better suited to a certain type of accelerator {[}6{]}, and choosing the wrong device results in slower and more energy-intensive computation {[}7{]}.
Thus accurate performance prediction is critical to making optimal scheduling decisions in a heterogeneous supercomputing environment.

The Architecture-Independent Workload Characterization (AIWC) tool {[}8{]} was previously introduced in order to collect architecture-independent features of OpenCL application workload.
AIWC operates on OpenCL kernels by simulating an OpenCL device and performing instrumentation to collect various features to characterize parallelism, compute complexity, memory and control that are independent of the target execution architecture.
In this paper, we propose a model that employs the AIWC features to make accurate predictions over a range of current accelerators.
These features are used to build a model which accurately predicts the execution times of a previously unseen OpenCL code over the range of available devices.
The performance predictions from this model may serve as input to scheduling decisions on heterogeneous supercomputing systems.

A major benefit of this approach is that the developer need only instrument a kernel once and the AIWC metrics can be embedded as a comment in the kernel's source code or Standard Portable Intermediate Representation (SPIR).
A scheduler system could be augmented to use the performance model with very low overhead, since querying the model is computationally inexpensive.
The model need only be retrained when a new accelerator type is added.
The methodology to develop the model is outlined in the following sections.
All tools used are open source, and all code is available in the respective repositories: {[}9{]} and {[}10{]}.

\section{Related Work}\label{related-work}

Augonnet et al. {[}11{]} propose a task scheduling framework for efficiently issuing work between multiple heterogeneous accelerators on a per-node basis.
They focus on the dynamic scheduling of tasks while automating data transfers between processing units to better utilise GPU-based HPC systems.
Much of this work is placed on evaluating the scaling of two applications over multiple nodes -- each of which are comprised of many GPUs.
Unfortunately, the presented methodology requires code to be rewritten using their MPI-like library.
OpenCL, by comparison, has been in use since 2008 and supports heterogeneous execution on most accelerator devices.
The algorithms presented to automate data movement should be reused for scheduling of OpenCL kernels to heterogeneous accelerator systems.

Existing works, {[}12{]}, {[}13{]}, {[}14{]} and {[}15{]}, have addressed heterogeneous distributed system scheduling and in particular the use of Directed Acyclic Graphs to track dependencies of high priority tasks.
Provided the parallelism of each dependency is expressed as OpenCL kernels, the model proposed here can be used to improve each of these scheduler algorithms by providing accurate estimates of execution time for each task for each potential accelerator on which the computation could be performed.

Our work is most closely related to efforts to enable low-cost performance estimates over a wide range of execution platforms.
One such approach uses partial execution, as introduced by Yang et al. {[}16{]}.
Here a short portion of a parallel code is executed and, since parallel codes are iterative behave predictably after the initial startup portion.
An important restriction for this approach is it requires execution on each of the accelerators for a given code, which may be complicated to achieve using common HPC scheduling systems.

An alternative performance prediction approach is given by Carrington et al. {[}17{]}.
Their solution generates two separate models each requiring two fundamental components: firstly, a machine profile of each system generated by running micro-benchmarks to probe simple performance attributes of each machine; and secondly, application signatures generated by instrumented runs which measure block information such as floating-point utilization and load/store unit usage of an application.
This is akin to our proposed solution using AIWC to generate each application signature and the generation of a random forest model to learn each machine profile.
However, in their method, no training takes place and the micro-benchmarks were developed with CPU memory hierarchy in mind, thus it is unsuited to a broader range of accelerator devices.
There are also many components and tools in use, for instance, network traffic is interpreted separately and requires the communication model to be developed from a different set of network performance capabilities, which needs more micro-benchmarks.
In comparison, our proposed solution uses a plugin to the Oclgrind tool, which is already widely used by OpenCL developers.

\section{Methodology}\label{methodology}

The AIWC tool {[}8{]} is a plugin to the Oclgrind {[}18{]} OpenCL device simulator, debugging and instrumentation tool.
AIWC simulates the execution of OpenCL kernels to collect architecture-independent features which characterize each code.
It operates on a restricted LLVM IR known as Standard Portable Intermediate Representation (SPIR) {[}19{]}, thereby simulating OpenCL kernel code in a hardware agnostic manner.
The AIWC metrics are shown in Table \ref{tbl:aiwc-metrics}.
We collected these metrics for a suite of benchmarks representative of scientific codes, which cover a wide spectrum of computation, communication and memory access patterns.
For each benchmark, we also collected detailed performance measurements on a varied set of compute devices.

\iftoggle{ACM-BUILD}{
\begin{table*}[t]
\caption{\textbf{AIWC} tool metrics. \label{tbl:aiwc-metrics}}
}{
\begin{table}[tb]
\caption{\textbf{AIWC} tool metrics. \label{tbl:aiwc-metrics}}
\begin{adjustbox}{max width=\textwidth}

}

\centering

\begin{tabular}{@{}cll@{}}
\toprule

{Type} & {Metric} & {Description}\\\hline

Compute & opcode & \# of unique opcodes required to cover 90\% of dynamic
instructions\\
Compute & Total Instruction Count & Total \# of instructions executed\\
Parallelism & Work-items & \# of work-items or threads executed\\
Parallelism & Total Barriers Hit & maximum \# of instructions executed until a barrier\\
Parallelism & Min ITB & minimum \# of instructions executed until a barrier\\
Parallelism & Max ITB & maximum \# of instructions executed until a barrier\\
Parallelism & Median ITB & median \# of instructions executed until a barrier\\
Parallelism & Max SIMD Width & maximum number of data items operated on during an instruction\\
Parallelism & Mean SIMD Width & mean number of data items operated on during an instruction\\
Parallelism & SD SIMD Width & standard deviation across the number of data items affected\\
Memory & Total Memory Footprint & \# of unique memory addresses accessed\\
Memory & 90\% Memory Footprint & \# of unique memory addresses that cover 90\% of memory
accesses\\
Memory & Global Memory Address Entropy & measure of the randomness of memory addresses\\
Memory & Local Memory Address Entropy & measure of the spatial locality of memory addresses\\
Control & Total Unique Branch Instructions & \# unique branch instructions\\
Control & 90\% Branch Instructions & \# unique branch instructions that cover 90\%
of branch instructions\\
Control & Yokota Branch Entropy & branch history entropy using Shannon's information entropy\\
Control & Average Linear Branch Entropy & branch history entropy score using the
average linear branch entropy\\
\hline
\end{tabular}

\iftoggle{ACM-BUILD}{
\end{table*}
}{
\end{adjustbox}
\end{table}
}

\subsection{Experimental Setup}\label{experimental-setup}

AIWC was used to characterize a variety of codes in the OpenDwarfs Extended (ODE) Benchmark Suite {[}20{]}, and the corresponding AIWC metrics were used as predictor variables in to fit a random forest regression model.
The metrics were generated over 4 problem sizes for each of the 11 applications -- and 37 computationally regions known as kernels in the OpenCL setting.
Response variables were collected following the same methodology outlined in {[}20{]} -- where the details for each of the applications is also presented.
Execution times were measured for at least 50 iterations and a total runtime of at least two seconds for each combination of device and benchmark.
Each application was run over 15 different accelerator devices, and are presented in Table \ref{tab:hardware}.
The L1 cache size should be read as having both an instruction cache and a data cache of the stated size.
For Nvidia GPUs, the L2 cache size reported is the size L2 cache per SM multiplied by the number of SMs.
For the Intel CPUs, Hyper-threading was enabled and the frequency governor was set to \texttt{performance}.

\begin{table*}[t]
\caption{Experimental hardware for generating runtime response data}
\centering
\begin{threeparttable}
    \centering
    \begin{tabular}{l|c|c|c|r|c|c|r|c}
        Name         & Vendor   & Type  & Series    & \multicolumn{1}{m{1cm}|}{\centering Core Count} & \multicolumn{1}{m{2.5cm}|}{\centering Clock Frequency (\si{\mega\hertz}) (min/max/turbo)}  &\multicolumn{1}{m{2.1cm}|}{\centering Cache (\SI{}{\kibi\byte}) (L1/L2/L3)} & \multicolumn{1}{m{.8cm}|}{\centering TDP (\SI{}{\watt})} &  \multicolumn{1}{m{1cm}}{\centering Launch  Date} \\ \hline
        Xeon E5-2697 v2  & Intel    & CPU   &Ivy Bridge & 24$\ast$ &1200/2700/3500 & 32/256/30720 & 130 & Q3 2013\\
        i7-6700K & Intel    & CPU   &Skylake & 8$\ast$ & 800/4000/4300 & 32/256/8192& 91 & Q3 2015\\
        i5-3550  & Intel    & CPU   & Ivy Bridge & 4$\ast$ & 1600/3380/3700 & 32/256/6144& 77 & Q2 2012\\
        Titan X & Nvidia & GPU & Pascal & 3584\textdagger & 1417/1531/-- & 48/2048/-- & 250 & Q3 2016\\
        GTX 1080 & Nvidia & GPU & Pascal & 2560\textdagger & 1607/1733/-- & 48/2048/-- & 180 & Q2 2016\\
        GTX 1080 Ti & Nvidia & GPU & Pascal & 3584\textdagger & 1480/1582/-- & 48/2048/-- & 250 & Q1 2017\\
        K20m & Nvidia & GPU & Kepler & 2496\textdagger & 706/--/-- & 64/1536/-- & 225 & Q4 2012\\
        K40m & Nvidia & GPU & Kepler & 2880\textdagger & 745/875/-- & 64/1536/-- & 235 & Q4 2013\\
        FirePro S9150 & AMD & GPU & Hawaii & 2816$\|$ & 900/--/-- & 16/1024/-- & 235 & Q3 2014\\
        HD 7970       & AMD & GPU & Tahiti & 2048$\|$ & 925/1010/-- & 16/768/-- & 250 & Q4 2011\\
        R9 290X       & AMD & GPU & Hawaii & 2816$\|$ & 1000/--/-- & 16/1024/--& 250 & Q3 2014\\
        R9 295x2      & AMD & GPU & Hawaii & 5632$\|$ & 1018/--/-- & 16/1024/--& 500 & Q2 2014\\
        R9 Fury X     & AMD & GPU & Fuji   & 4096$\|$ & 1050/--/-- & 16/2048/--& 273 & Q2 2015\\
        RX 480        & AMD & GPU & Polaris& 4096$\|$ & 1120/1266/-- & 16/2048/-- & 150 & Q2 2016\\
        Xeon Phi 7210 & Intel & MIC & KNL & 256\textdaggerdbl & 1300/1500/-- & 32/1024/-- & 215 & Q2 2016\\
    \end{tabular}
    \begin{tablenotes}
    \item [$\ast$] HyperThreaded cores
    \item [\textdagger] CUDA cores
    \item [$\|$] Stream processors
    \item [\textdaggerdbl] Each physical core has 4 hardware threads per core, thus 64 cores
    \end{tablenotes}
\end{threeparttable}
\label{tab:hardware}
\end{table*}

\subsection{Constructing the Performance Model}\label{constructing-the-performance-model}

The R programming language was used to analyse the data, construct the model and analyse the results.
In particular, the ranger package by Wright and Ziegler {[}21{]} was used for the development of the regression model.
The ranger package provides computationally efficient implementations of the Random Forest model {[}22{]} which performs recursive partitioning of high dimensional data.

The ranger function accepts three main parameters, each of which influences the fit of the model to the data.
In optimizing the model, we searched over a range of values for each parameter including:

\begin{itemize}
\tightlist
\item
  num.trees, the number of trees grown in the random forest: over the range of \(10 - 10,000 \text{ by } 500\)
\item
  mtry, the number of features tried to possibly split within each node: ranges from \(1 - 34\), where \(34\) is the maximum number of input features available from AIWC,
\item
  min.node.size, the minimal node size per tree: ranges from \(1 - 50\), where \(50\) is the number of observations per sample.
\end{itemize}

Given the size of the data set, it was not computationally viable to perform an exhaustive search of the entire 3-dimensional range of parameters.
Auto-tuning to determine the suitability of these parameters has been performed by Lie\ss~et al. {[}23{]} to determine the optimal value of mtry given a fixed num.trees.
Instead, to enable an efficient search of all variables at once, we used Flexible Global Optimization with Simulated-Annealing, in particular, the variant found in the R package \textit{optimization} by Husmann, Lange and Spiegel {[}24{]}.
The simulated-annealing method both reduces the risk of getting trapped in a local minimum and is able to deal with irregular and complex parameter spaces as well as with non-continuous and sophisticated loss functions.
In this setting, it is desirable to minimise the out-of-bag prediction error of the resultant fitted model, by simultaneously changing the parameters (num.trees, mtry and min.node.size).
The \textit{optim\_sa} function allows defining the search space of interest, a starting position, the magnitude of the steps according to the relative change in temperature and the wrapper around the ranger function (which parses the 3 parameters and returns a cost function --- the predicted error).
It allows for an approximate global minimum to be detected with significantly fewer iterations than an exhaustive grid search.

\begin{figure}[t]
\centering
\iftoggle{ACM-BUILD}{
\includegraphics[width=0.9\columnwidth]{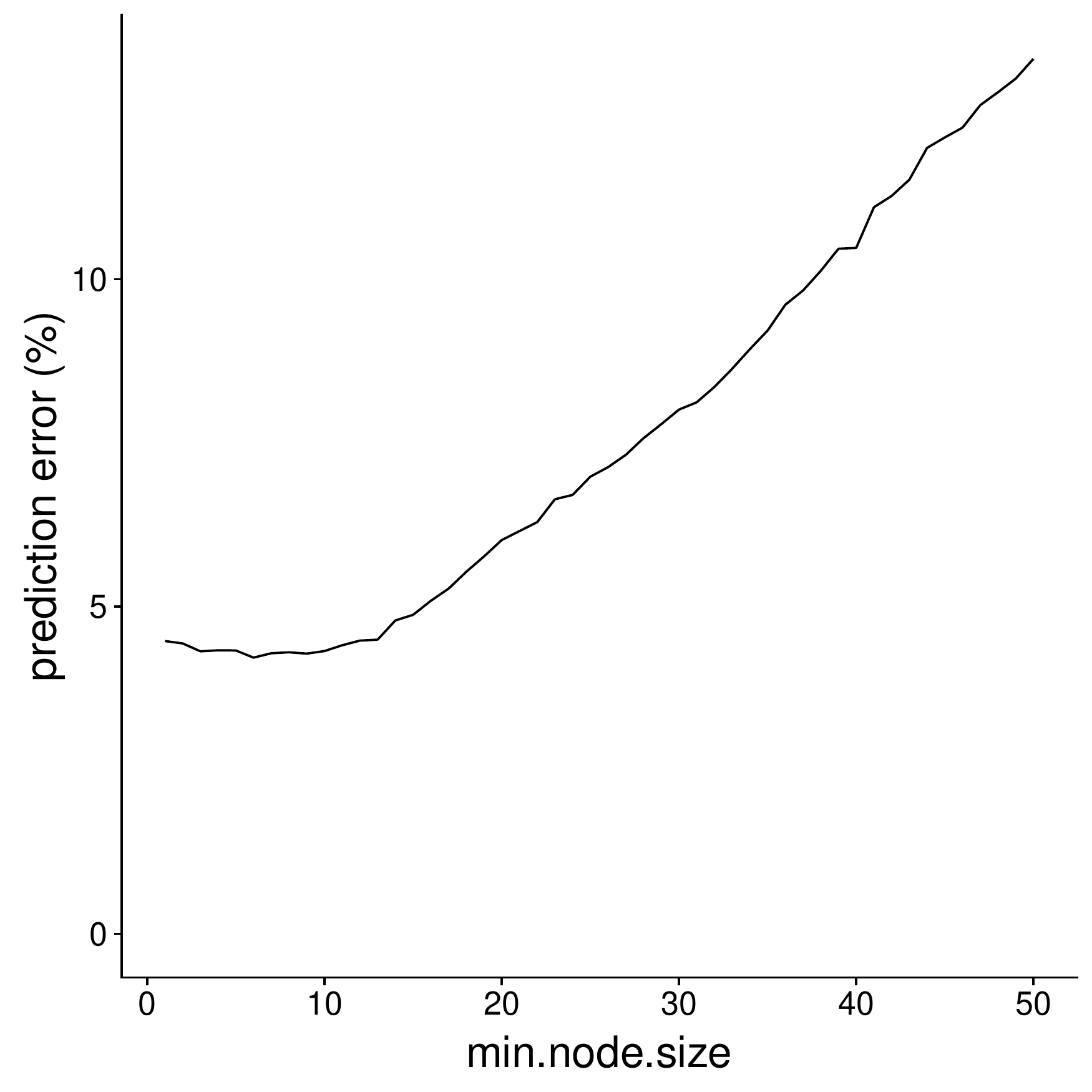}
}{
\includegraphics[width=0.45\textwidth]{./figure/full-variation-in-min-node-size-1.pdf}
}
\caption{\label{fig:variation-in-min-node-size}Full coverage of min.node.size with fixed tuning parameters: num.trees = 300 and mtry = 30.}
\end{figure}

\begin{figure}[t]
\centering
\iftoggle{ACM-BUILD}{
\includegraphics[width=0.9\columnwidth]{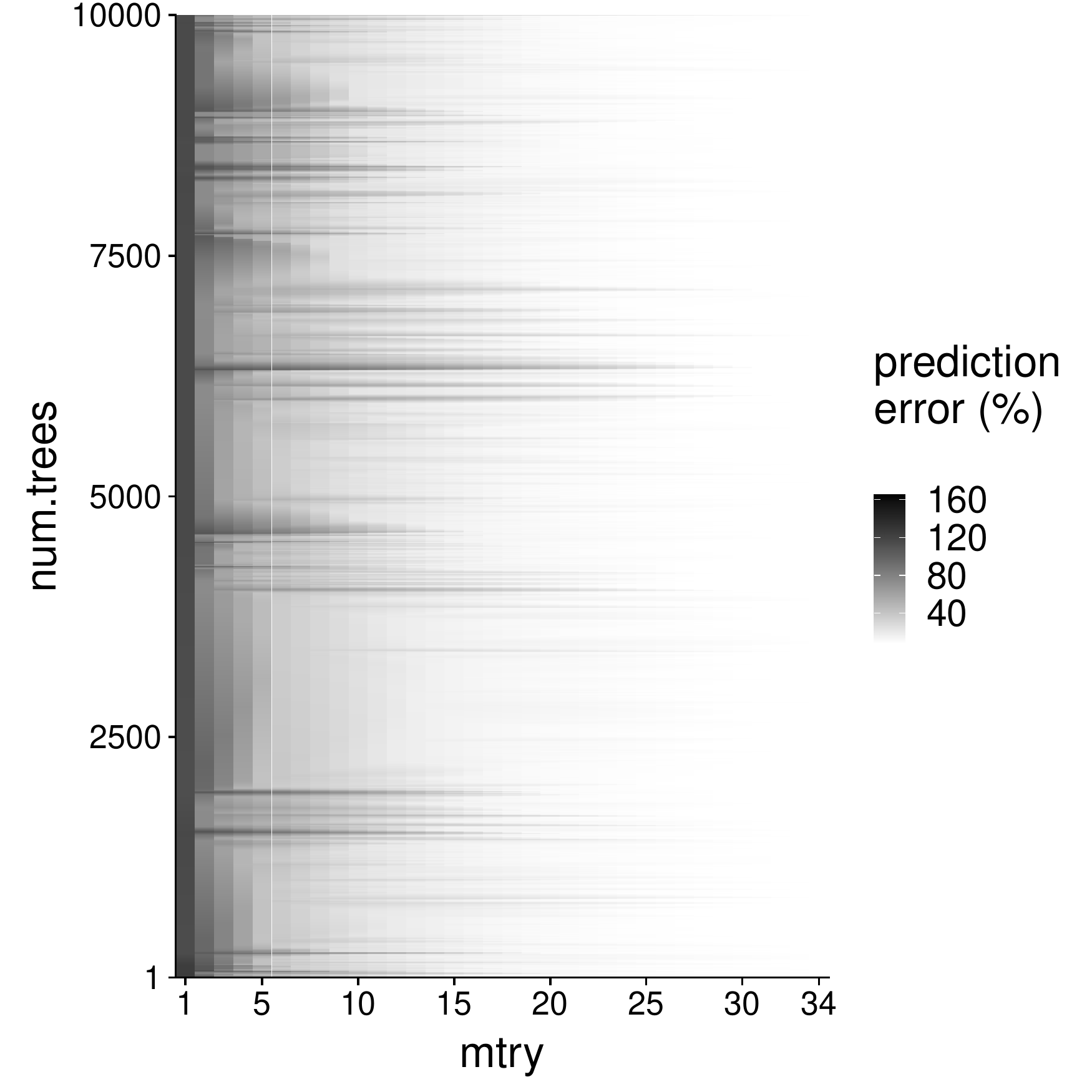}
}{
\includegraphics[width=0.45\textwidth]{./figure/full-scan-random-sampled-heatmap-1.pdf}
}
\caption{\label{fig:full-scan-random-sampled-heatmap}Full coverage of num.trees and mtry tuning parameters with min.node.size fixed at 9.}
\end{figure}

Figure \ref{fig:variation-in-min-node-size} shows the relationship between out-of-bag prediction error and min.node.size, with the num.trees = 300 and mtry = 30 parameters fixed.
In general, the min.node.size has the smallest prediction error for values less than 15 and variation in prediction error is similar throughout this range.
As such, the selection to fix min.node.size = 9 was made to reduce the search-space in the remainder of the tuning work.
We assume conditional (relative) independence between min.node.size and the other variables.

Figure \ref{fig:full-scan-random-sampled-heatmap} shows how the prediction error of the random-forest ranger model changes over a wide range of values for the two remaining tuning parameters, mtry and num.trees.
Full coverage was achieved by selecting starting locations in each of the 4 outer-most points of the search space, along with 8 random internal points --- to avoid missing out on some critical internal structure.
For each combination of parameter values, the \textit{optim\_sa} function was allowed to execute until a global minimum was found.
At each step of optimization a full trace was collected, where all parameters and the corresponding out-of-bag prediction error value were logged to a file.
This file was finally loaded, the points interpolated using the R package akima, without extrapolation between points, using the mean values for duplication between points.
The generated heatmap is shown in Figure \ref{fig:full-scan-random-sampled-heatmap}.

A lower out-of-bag prediction error is better.
For values of mtry above 25, there is good model fit irrespective of the number of trees.
For lower values of mtry, fit varies significantly with different values of num.trees.
The worst fit was for a model with a value of 1 num.trees, and 1 for mtry, which had the highest out-of-bag prediction error at 194\%.
In general, the average prediction error across all choices of parameters is very low at 16\%.
Given these results, the final ranger model should use a small value for num.trees and a large value for mtry, with the added benefit that such a model can be computed faster given a smaller number of trees.

\subsection{\texorpdfstring{Choosing Model Parameters \label{sec:choosing-model-parameters}}{Choosing Model Parameters }}\label{choosing-model-parameters}

The selected model should be able to accurately predict execution times for a previously unseen kernel over the full range of accelerators.
To show this, the model must not be over-fitted, that is to say, the random forest model parameters should not be tuned to the particular set of kernels in the training data, but should generate equally good fits if trained on any other reasonable selection of kernels.

We evaluated how robust the selection of model parameters is to the choice of kernel by repeatedly retraining the model on a set of kernels, each time removing a different kernel.
The procedure used is presented in Algorithm \ref{alg:kernel-omission}.
For each selection of kernels, \textit{optima\_sa} was run from the same starting location -- num.trees=500, mtry=32 -- and the final optimal values were recorded. min.node.size was fixed at 9.

The optimal -- and final -- parameters for each omitted kernel are presented in Table \ref{tab:optimal-tuning-parameters}.
Regardless of which kernel is omitted, the R-squared values -- or explained variance -- is very high at 0.99, indicating a good model fit.
The optimal parameters are very similar regardless of which kernel was omitted.
As such, the median value of each of the parameters was selected for the final model: num.trees = 505, mtry = 30 and min.node.size = 9.
These parameters were used for all further model training.

\begingroup
\removelatexerror

\begin{algorithm}[t]

    \For{each unique kernel}{
        construct a full data frame with all but the current kernel\;
        run optimization \textit{optim\_sa} with the full data frame at selected starting location\;
        record the final optimal parameters
    }
    \caption{\label{alg:kernel-omission}Find the suitability of the optimal parameters for random forest models for future kernels}
\end{algorithm}

\endgroup

\iftoggle{ACM-BUILD}{
\begin{table}[t]
\caption{Optimal tuning parameters from the same starting location for all models omitting each individual kernel.\label{tab:optimal-tuning-parameters}}
}{
\begin{table}[tb]
\begin{adjustbox}{max width=\textwidth}
\caption{Optimal tuning parameters from the same starting location for all models omitting each individual kernel.\label{tab:optimal-tuning-parameters}}
}

\centering

\begin{tabularx}{\columnwidth}{@{}cccc@{}}
\toprule

{Kernel omitted} & {num.trees} & {mtry} & \multicolumn{1}{m{1cm}}{\centering prediction error (\%)}\\\hline

invert\_mapping & 521 & 31 & 4.3\\
kmeansPoint & 511 & 30 & 4.1\\
lud\_diagonal & 527 & 29 & 4.4\\
lud\_internal & 488 & 31 & 4.5\\
lud\_perimeter & 480 & 31 & 4.4\\
csr & 507 & 30 & 4.4\\
fftRadix16Kernel & 484 & 29 & 4.4\\
fftRadix8Kernel & 529 & 34 & 4.3\\
fftRadix4Kernel & 463 & 30 & 4.2\\
fftRadix2Kernel & 443 & 28 & 4.4\\
calc\_potential\_single\_step & 502 & 24 & 4.8\\
c\_CopySrcToComponents & 529 & 31 & 4.1\\
cl\_fdwt53Kernel & 499 & 26 & 4.7\\
srad\_cuda\_1 & 504 & 32 & 4.7\\
srad\_cuda\_2 & 500 & 29 & 4.6\\
kernel1 & 536 & 30 & 4.5\\
kernel2 & 469 & 31 & 4.6\\
acc\_b\_dev & 576 & 28 & 4.4\\
calc\_alpha\_dev & 469 & 30 & 4.3\\
calc\_beta\_dev & 498 & 30 & 4.3\\
calc\_gamma\_dev & 517 & 28 & 4.4\\
calc\_xi\_dev & 439 & 33 & 4.3\\
est\_a\_dev & 524 & 30 & 4.2\\
est\_b\_dev & 533 & 28 & 4.3\\
est\_pi\_dev & 450 & 31 & 4.3\\
init\_alpha\_dev & 558 & 32 & 2.6\\
init\_beta\_dev & 467 & 30 & 4.1\\
init\_ones\_dev & 566 & 32 & 4.1\\
mvm\_non\_kernel\_naive & 514 & 30 & 4.3\\
mvm\_trans\_kernel\_naive & 449 & 32 & 4.4\\
scale\_a\_dev & 508 & 31 & 4.3\\
scale\_alpha\_dev & 530 & 30 & 3.8\\
scale\_b\_dev & 565 & 31 & 4.2\\
s\_dot\_kernel\_naive & 509 & 30 & 4.5\\
needle\_opencl\_shared\_1 & 499 & 30 & 4.4\\
needle\_opencl\_shared\_2 & 504 & 29 & 4.5\\
crc32\_slice8 & 511 & 29 & 4.3\\

\hline

\end{tabularx}

\iftoggle{ACM-BUILD}{
\end{table}
}{
\end{adjustbox}
\end{table}
}

\subsection{\texorpdfstring{Performance Improvement with Increased Training Data \label{sec:finding-the-critical-number-of-kernels}}{Performance Improvement with Increased Training Data }}\label{performance-improvement-with-increased-training-data}

For a model to be useful in predicting execution times for previously unseen kernels, it needs to be trained on a representative sample of kernels i.e.~a sample that provides good coverage of the AIWC feature space of all possible application kernels.

We measured how model fit improves with the number of kernels used in training, following the method presented in Algorithm \ref{alg:rmse-per-kernel-count}.
The set of unique kernels available during model development is denoted by \(k\) (37 kernels in this study), \(s\) is the maximum number of sample models (including different combinations of kernels) to evaluate for each number of kernels 1..\(|k|\), \(\phi\) is a data frame of the combined AIWC feature-space with measured runtime results.
The parameters to the random forest model were fixed at num.trees = 505, mtry = 30 and min.node.size = 9, according to the methodology in Section \ref{sec:choosing-model-parameters}.

\begingroup
\removelatexerror
\newcommand{\isep}{\mathrel{{.}\,{.}}\nobreak}

\begin{algorithm}[t]

    $s \gets 500$\\
    $k \gets $unique(kernel)\\
    \For{$i \gets 1 $\textbf{to} length($k$)}{
        $v_p \gets [ ]$\\
        $v_m \gets [ ]$\\
        \For{$j \gets 1$ \textbf{to} $s$}{
            $x \gets $shuffle($k$)\\
            $y \gets x[1 \isep i]$\\
            \textbf{training data} $\gets$ subset($\phi$, kernel $== y$) \\
            \textbf{test data} $\gets$ subset($\phi$, kernel $!= y$) \\
            discard variables unavailable during real-world training from \textbf{training data} e.g. size, application, kernel name and measured total application time\\
            build ranger model $r$ using \textbf{training data} \\
            generate prediction responses $p$ from $r$ using \textbf{test data}\\
            append predicted execution times $p$ to $v_p$\\
            append measured execution times from \textbf{test data} to $v_m$\\
        }
        compute the mean absolute error $e$ from vector of $p$ relative to vector $m$\\
        store($e$)\\
    }

    \caption{\label{alg:rmse-per-kernel-count}Compute average fit of random forest models trained on different numbers of kernels.}

\end{algorithm}

\endgroup

\begin{figure}[htbp]
\centering
\iftoggle{ACM-BUILD}{
\includegraphics[width=0.9\columnwidth]{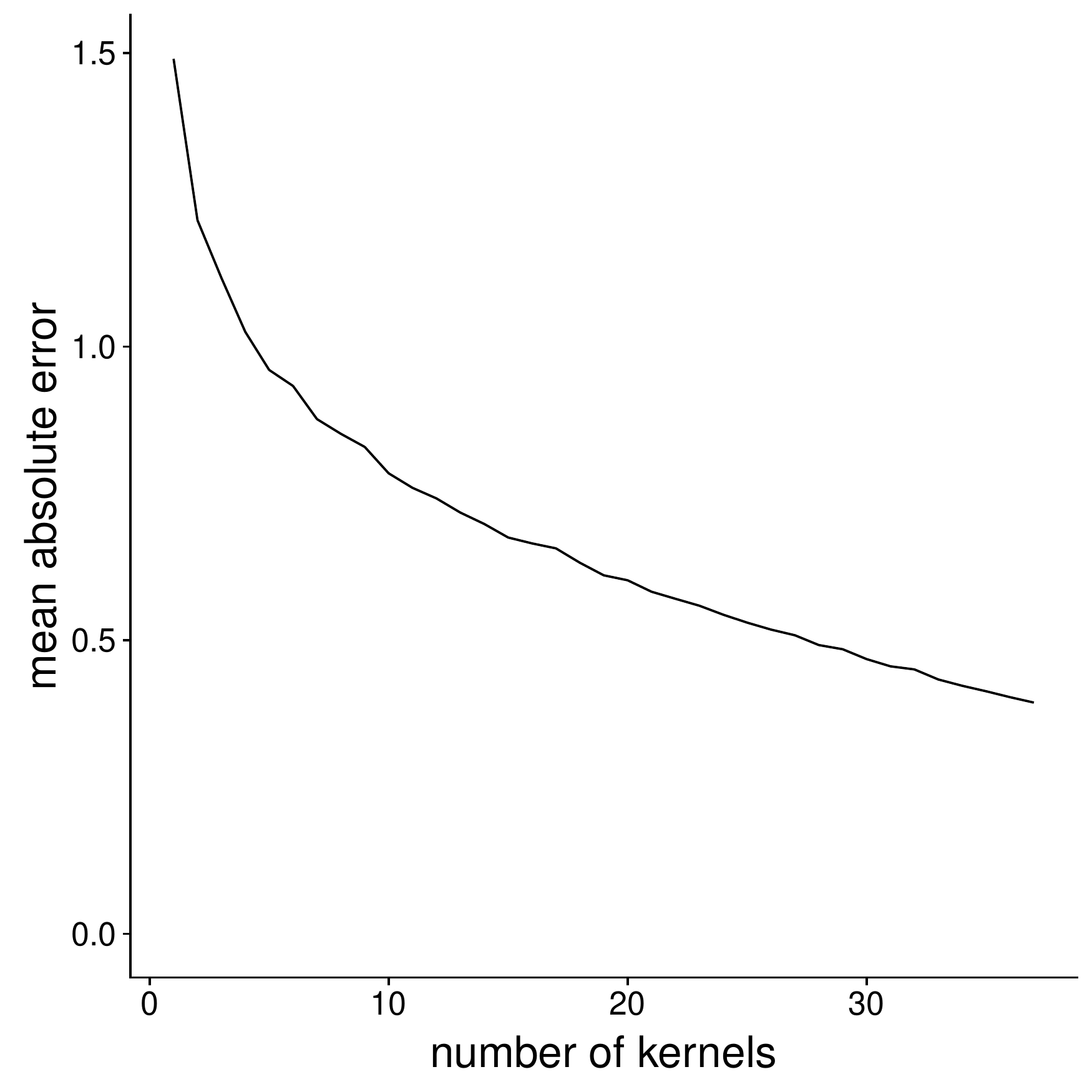}
}{
\includegraphics[width=0.45\textwidth]{./figure/rmse_vs_kernel_count-1.pdf}
}
\caption{\label{fig:rmse-vs-kernel-count}Prediction error across all benchmarks for models trained with varying numbers of kernels.}
\end{figure}

The results presented in Figure \ref{fig:rmse-vs-kernel-count} show the mean absolute error of models trained on varying numbers of kernels.
As expected, the model fit improves with increasing number of kernels.
In particular, larger improvements occur with each new kernel early in the series and tapers off as a new kernel is added to an already large number of kernels.
The gradient is still significant until the largest number of samples examined (\(k=37\)) suggesting that the model could benefit from additional training data.
However, the model proposed is a proof of concept and suggests that a general purpose model is attainable and may not require many more kernels.

\section{Evaluation}\label{evaluation}

\begin{figure}[htbp]
\centering
\iftoggle{ACM-BUILD}{
\includegraphics[width=0.9\columnwidth]{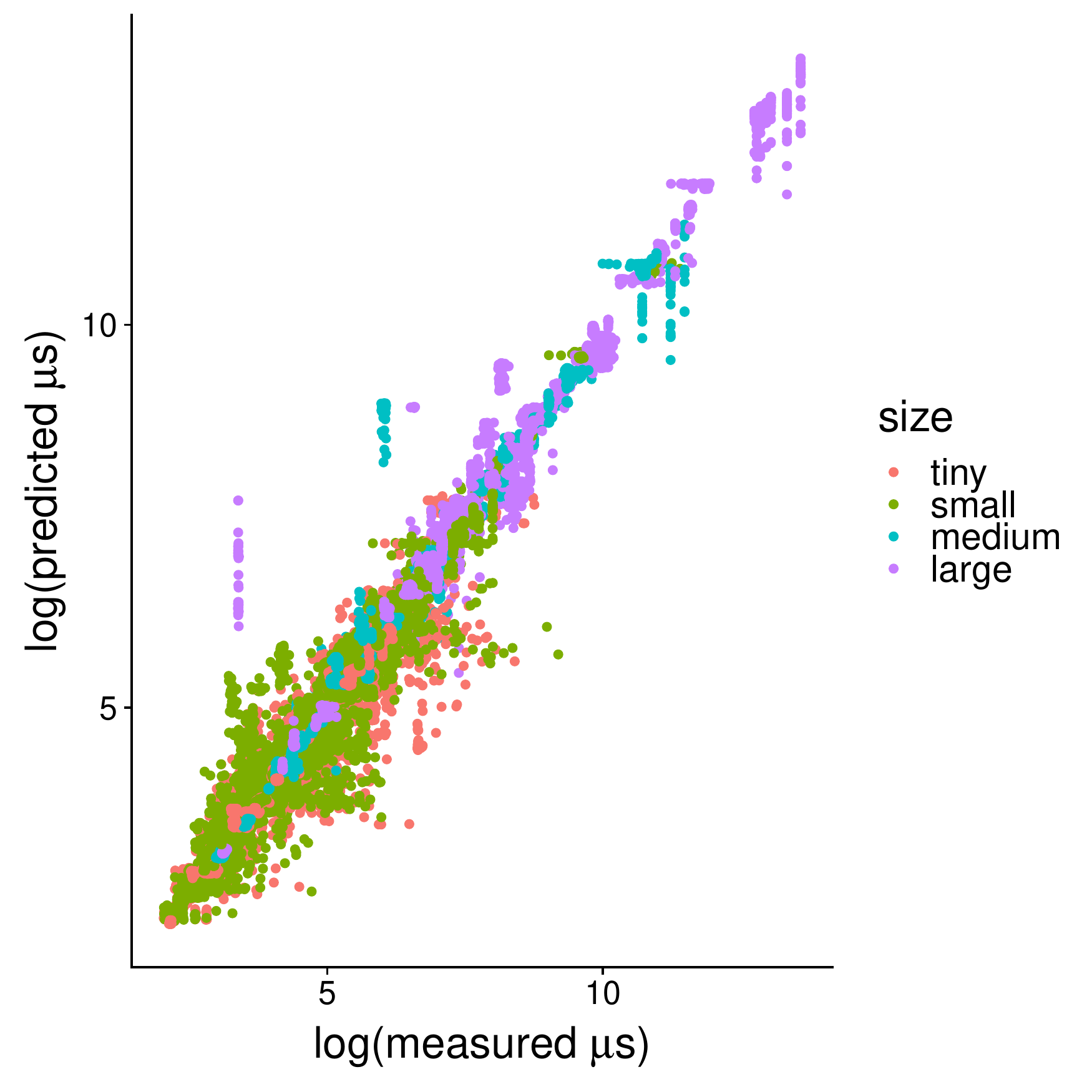}
}{
\includegraphics[width=0.45\textwidth]{./figure/actual-vs-predicted-size-plot-1.pdf}
}
\caption{\label{fig:selected-model-actual-vs-predicted-times}Predicted vs. measured execution time for all kernels}
\end{figure}

Figure \ref{fig:selected-model-actual-vs-predicted-times} presents the measured kernel execution times against the predicted execution times from the trained model.
Each point represents a single combination of kernel and problem size.
The plot shows a strong linear correlation indicating a good model fit.
Under-predictions typically occur on four kernels over the medium and large problem sizes, while over-predictions occur on the tiny and small problem sizes.
However, these outliers are visually over-represented in this figure as the final mean absolute error is low, at \textasciitilde{}0.11.

\section{Making Predictions}\label{making-predictions}

In this section, we examine differences in accuracy of predicted execution times between different kernels, which is of importance if the predictions are to be used in a scheduling setting.

\begin{figure*}
\centering
\iftoggle{ACM-BUILD}{
\includegraphics[width=0.95\linewidth]{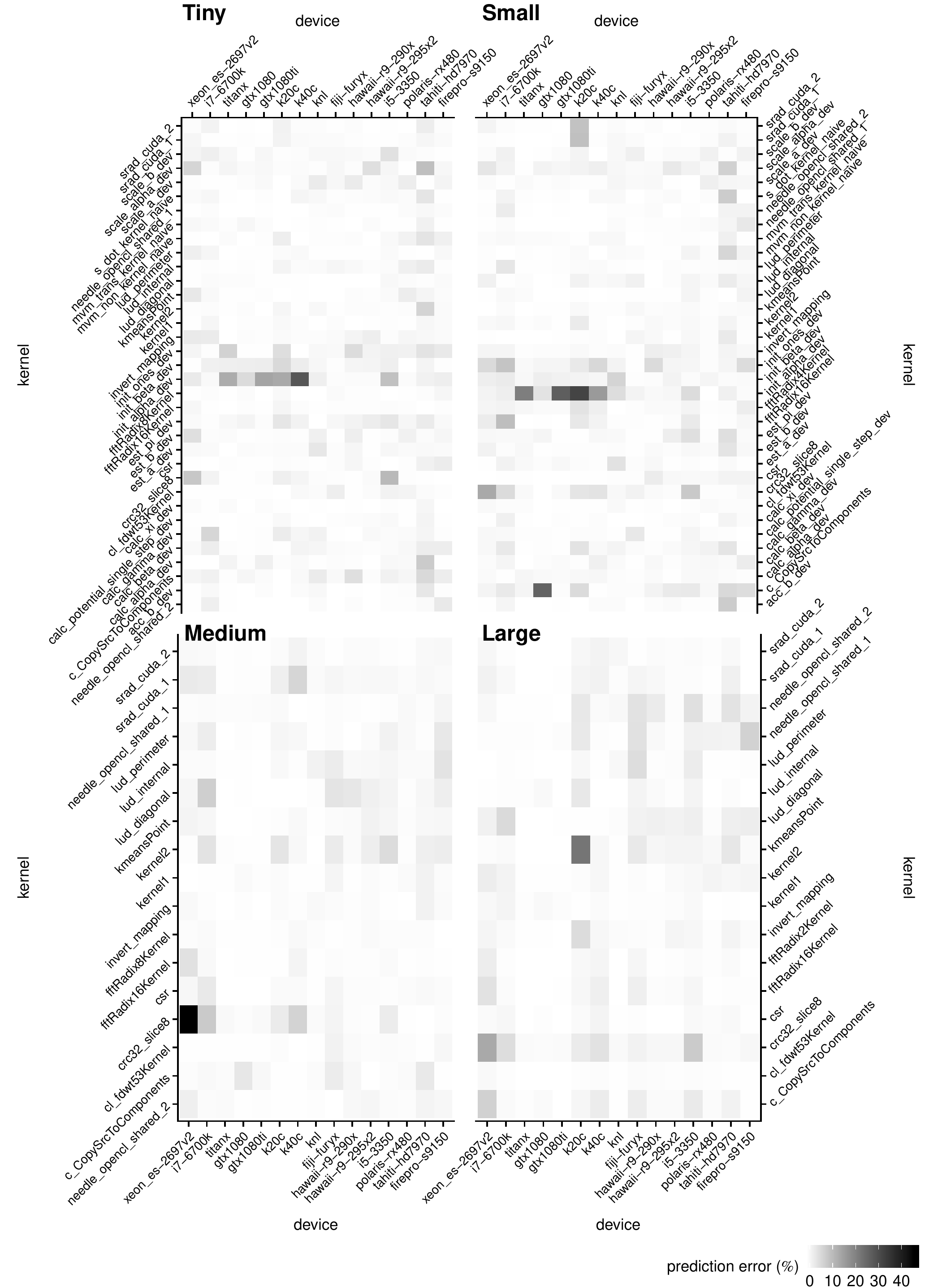}
}{
\includegraphics[width=0.95\textwidth,height=0.95\textheight,keepaspectratio]{./figure/predictive-heatmap-percentage-1.pdf}
}
\caption{\label{fig:predictive-heatmap-percentage}Error in predicted execution time for each kernel invocation over four problem sizes}
\end{figure*}

The four heat maps presented in Figure \ref{fig:predictive-heatmap-percentage} show the difference between mean predicted and measured kernel execution times as a percentage of the measured time.
Thus, they depict the relative error in prediction -- lighter indicates a smaller error.
Four different problem sizes are presented: tiny in the top-left, small in the top-right, medium bottom-left, large bottom-right.

In general, we see highly accurate predictions which on average differ from the measured experimental run-times by 1.1\%, which correspond to actual execution time mispredictions of 8 \(\mu s\) to 1 secs according to problem size.

The \texttt{init\_alpha\_dev} kernel is the worst predicted kernel over both the tiny and small problem sizes, with mean misprediction at 7.3\%.
However, this kernel is only run once per application run -- it is used in the initialization of the Hidden Markov Model -- and as such there are fewer response variables available for model training.

\section{The benefits of this approach}\label{the-benefits-of-this-approach}

\begin{figure*}
\centering
\iftoggle{ACM-BUILD}{
\includegraphics[width=\linewidth]{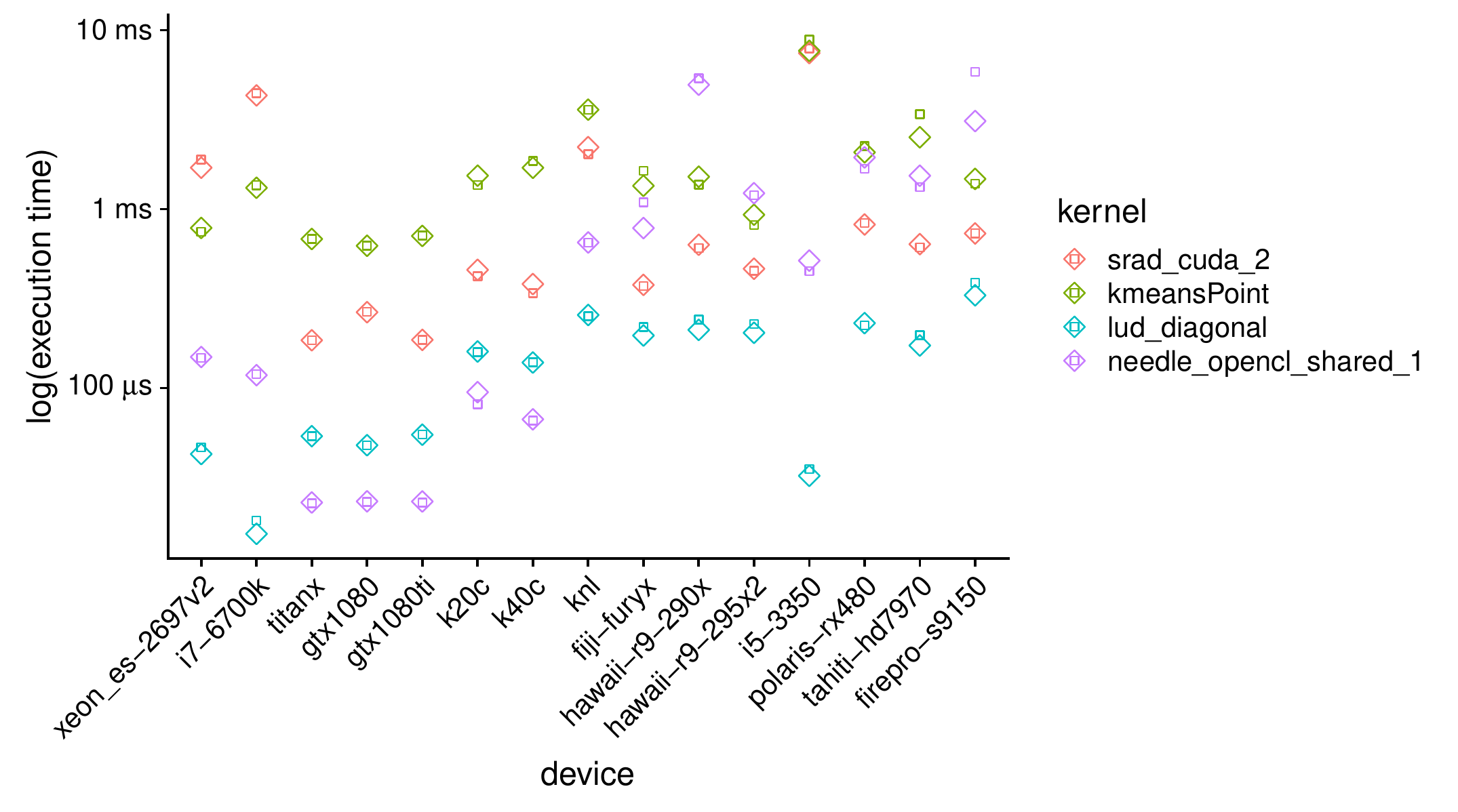}
}{
\includegraphics[width=0.95\textwidth,height=0.95\textheight,keepaspectratio]{./figure/large-predicted-vs-measured-1.pdf}
}
\caption{\label{fig:large-predicted-vs-measured}Mean measured kernel execution times compared against mean predicted kernel execution times to perform a selection of kernels on large problem sizes across 15 accelerator devices.}
\end{figure*}

To demonstrate the utility of the trained model to guide scheduling choices, we focus on the accuracy of performance time prediction of individual kernels over all devices.
The model performance in terms of real execution times is presented for four randomly selected kernels in Figure \ref{fig:large-predicted-vs-measured}.
The shape denotes the type of execution time data point, a square indicates the mean measured time, and the diamond indicates the mean predicted time.
Thus, a perfect prediction occurs where the measured time -- square -- fits perfectly within the predicted -- diamond -- as seen in the legend.

The purpose of showing these results is to highlight the setting in which they could be used -- on the supercomputing node.
In this instance, it is expected a node to be composed of any combination of the 15 devices presented in the Figure \ref{fig:large-predicted-vs-measured}.
Thus, to be able to advise a scheduler which device to use to execute a kernel, the model must be able to correctly predict on which of a given pair of devices the kernel will run fastest.
For any selected pair of devices, if the relative ordering of the measured and predicted execution times is different, the scheduler would choose the wrong device.
In almost all cases, the relative order is preserved using our model.
In other words, our model will correctly predict the fastest device in all cases -- with one exception, the \texttt{kmeansPoint} kernel.
For this kernel, the predicted time of the fiji-furyx is lower than the hawaii-r9-290x, however the measured times between the two shows the furyx completing the task in a shorter time.
For all other device pairs, the relative order for the \texttt{kmeansPoint} kernel is correct.
Additionally, the \texttt{lud\_diagonal} kernel suffers from systematic under-prediction of execution times on AMD GPU devices, however the relative ordering is still correct.
As such, the proposed model provides sufficiently accurate execution time predictions to be useful for scheduling to heterogeneous compute devices on supercomputers.

\section{Conclusions and Future Work}\label{conclusions-and-future-work}

A highly accurate model has been presented that is capable of predicting execution times of OpenCL kernels on specific devices based on the computational characteristics captured by the AIWC tool.
A real-world scheduler could be developed based on the accuracy of the presented model.

We do not suppose that we have used a fully representative suite of kernels, however, we have shown that this approach can be used in the supercomputer accelerator scheduling setting, and the model can be extended/augmented with additional training kernels using the methodology presented in this paper.

We expect that a similar model could be constructed to predict energy or power consumption, where the response variable can be directly swapped for an energy consumption metric -- such as joules -- instead of execution time.
However, we have not yet collected the energy measurements required to construct such a model.
Finally, we show the predictions made are accurate enough to inform scheduling decisions.


\begin{thebibliography}{00}
\hypertarget{refs}{}
\bibitem{b1}\hypertarget{ref-declerck2016cori}{} T. Declerck \emph{et al.}, ``Cori - a system to support data-intensive computing,'' \emph{Proceedings of the Cray User Group}, p. 8, 2016.
\bibitem{b2}\hypertarget{ref-morgan_2016}{} T. Morgan, ``NVLink takes GPU acceleration to the next level,'' \emph{The Next Platform}, May 2016.
\bibitem{b3}\hypertarget{ref-morgan_2017}{} T. Morgan, ``The Power9 rollout begins with Summit and Sierra supercomputers,'' \emph{The Next Platform}, Sep. 2017.
\bibitem{b4}\hypertarget{ref-morgan_2016_postk}{} T. Morgan, ``Inside Japan's future exascale ARM supercomputer,'' \emph{The Next Platform}. Stackhouse Publishing Inc., Jun-2016.
\bibitem{b5}\hypertarget{ref-feldman_2017_isambard}{} M. Feldman, ``Cray to deliver ARM-powered supercomputer to UK consortium,'' \emph{TOP500 Supercomputer Sites}, Jan. 2017.
\bibitem{b6}\hypertarget{ref-che2008accelerating}{} S. Che, J. Li, J. W. Sheaffer, K. Skadron, and J. Lach, ``Accelerating compute-intensive applications with gpus and fpgas,'' in \emph{Application specific processors, 2008. sasp 2008. symposium on}, 2008, pp. 101--107.
\bibitem{b7}\hypertarget{ref-yildirim2012single}{} M. B. Yildirim and G. Mouzon, ``Single-machine sustainable production planning to minimize total energy consumption and total completion time using a multiple objective genetic algorithm,'' \emph{IEEE transactions on engineering management}, vol. 59, no. 4, pp. 585--597, 2012.
\bibitem{b8}\hypertarget{ref-aiwc2018}{} B. Johnston and J. Milthorpe, ``AIWC: OpenCL based Architecture Independent Workload Characterisation,'' \emph{ArXiv e-prints}, May 2018.
\bibitem{b9}\hypertarget{ref-johnston2017}{} B. Johnston, ``OpenDwarfs,'' \emph{GitHub repository}. \url{https://github.com/BeauJoh/OpenDwarfs}; GitHub, 2017.
\bibitem{b10}\hypertarget{ref-beau_johnston_2017_1134175}{} B. Johnston \emph{et al.}, ``BeauJoh/Oclgrind: Adding AIWC -- An Architecture Independent Workload Characterisation Plugin.'' Dec-2017.
\bibitem{b11}\hypertarget{ref-augonnet2010data}{} C. Augonnet, J. Clet-Ortega, S. Thibault, and R. Namyst, ``Data-aware task scheduling on multi-accelerator based platforms,'' in \emph{IEEE international conference on parallel and distributed systems (ICPADS)}, 2010, pp. 291--298.
\bibitem{b12}\hypertarget{ref-topcuoglu1999task}{} H. Topcuoglu, S. Hariri, and M.-Y. Wu, ``Task scheduling algorithms for heterogeneous processors,'' in \emph{Heterogeneous computing workshop (HCW)}, 1999, pp. 3--14.
\bibitem{b13}\hypertarget{ref-bajaj2004improving}{} R. Bajaj and D. P. Agrawal, ``Improving scheduling of tasks in a heterogeneous environment,'' \emph{IEEE Transactions on Parallel and Distributed Systems}, vol. 15, no. 2, pp. 107--118, 2004.
\bibitem{b14}\hypertarget{ref-xiaoyong2011novel}{} T. Xiaoyong, K. Li, Z. Zeng, and B. Veeravalli, ``A novel security-driven scheduling algorithm for precedence-constrained tasks in heterogeneous distributed systems,'' \emph{IEEE Transactions on Computers}, vol. 60, no. 7, pp. 1017--1029, 2011.
\bibitem{b15}\hypertarget{ref-sinnen2004list}{} O. Sinnen and L. Sousa, ``List scheduling: Extension for contention awareness and evaluation of node priorities for heterogeneous cluster architectures,'' \emph{Parallel Computing}, vol. 30, no. 1, pp. 81--101, 2004.
\bibitem{b16}\hypertarget{ref-yang2005cross}{} L. T. Yang, X. Ma, and F. Mueller, ``Cross-platform performance prediction of parallel applications using partial execution,'' in \emph{Proceedings of the 2005 ACM/IEEE conference on Supercomputing}, 2005, p. 40.
\bibitem{b17}\hypertarget{ref-carrington2006performance}{} L. Carrington, A. Snavely, and N. Wolter, ``A performance prediction framework for scientific applications,'' \emph{Future Generation Computer Systems}, vol. 22, no. 3, pp. 336--346, 2006.
\bibitem{b18}\hypertarget{ref-price:15}{} J. Price and S. McIntosh-Smith, ``Oclgrind: An extensible OpenCL device simulator,'' in \emph{Proceedings of the 3rd international workshop on OpenCL}, 2015, p. 12.
\bibitem{b19}\hypertarget{ref-kessenich2015}{} J. Kessenich, ``A Khronos-Defined Intermediate Language for Native Representation of Graphical Shaders and Compute Kernels.'' 2015.
\bibitem{b20}\hypertarget{ref-johnston17opendwarfs}{} B. Johnston and J. Milthorpe, ``Dwarfs on Accelerators: Enhancing OpenCL Benchmarking for Heterogeneous Computing Architectures,'' \emph{ArXiv e-prints}, May 2018.
\bibitem{b21}\hypertarget{ref-JSSv077i01}{} M. Wright and A. Ziegler, ``ranger: A Fast Implementation of Random Forests for High Dimensional Data in C++ and R,'' \emph{Journal of Statistical Software, Articles}, vol. 77, no. 1, pp. 1--17, 2017.
\bibitem{b22}\hypertarget{ref-breiman2001random}{} L. Breiman, ``Random forests,'' \emph{Machine learning}, vol. 45, no. 1, pp. 5--32, 2001.
\bibitem{b23}\hypertarget{ref-liess2014sloping}{} M. Ließ, M. Hitziger, and B. Huwe, ``The sloping mire soil-landscape of southern Ecuador: Influence of predictor resolution and model tuning on random forest predictions,'' \emph{Applied and environmental soil science}, vol. 2014, 2014.
\bibitem{b24}\hypertarget{ref-husmannr}{} K. Husmann, A. Lange, and E. Spiegel, ``The R package optimization: Flexible global optimization with simulated-annealing,'' 2017.
\end{thebibliography}
\end{document}